\documentclass[%
 reprint,
superscriptaddress,
 amsmath,amssymb,
 aps,
]{revtex4-2}
\usepackage{mhchem}
\usepackage{graphicx}
\usepackage{dcolumn}
\usepackage{bm}

\begin{document}

\preprint{APS/123-QED}

\title{Cavity Tuning of the CDW--Superconductivity Interplay in a Kagome Metal}

\author{Lan-Ting Shi}
\thanks{These authors contributed equally to this work.}
\affiliation{Tsientang Institute for advaced study, Zhejiang 310024, China}
\affiliation{Songshan Lake Materials Laboratory, Guangdong 523808, China}

\author{I-Te Lu}
\thanks{These authors contributed equally to this work.}
\affiliation{Max Planck Institute for the Structure and Dynamics of Matter and Center for Free-Electron Laser Science, Luruper Chaussee 149, Hamburg 22761, Germany}

\author{Xin Wang}
\affiliation{Songshan Lake Materials Laboratory, Guangdong 523808, China}
\affiliation{Department of Materials Science and Engineering,
City University of Hong Kong, Kowloon 999077, Hong Kong}

\author{Dongbin Shin}
\email{dshin@gist.ac.kr}
\affiliation{Max Planck Institute for the Structure and Dynamics of Matter and Center for Free-Electron Laser Science, Luruper Chaussee 149, Hamburg 22761, Germany}
\affiliation{Department of Physics and Photon Science, Gwangju Institute of Science and Technology (GIST), Gwangju 61005, Republic of Korea}

\author{Lede Xian}
\email{ldxian@tias.ac.cn}
\affiliation{Tsientang Institute for advaced study, Zhejiang 310024, China}
\affiliation{Max Planck Institute for the Structure and Dynamics of Matter and Center for Free-Electron Laser Science, Luruper Chaussee 149, Hamburg 22761, Germany}

\author{Angel Rubio}
\affiliation{Max Planck Institute for the Structure and Dynamics of Matter and Center for Free-Electron Laser Science, Luruper Chaussee 149, Hamburg 22761, Germany}
\affiliation{Initiative for Computational Catalysis (ICC) and Center for Computational Quantum Physics (CCQ), The Flatiron Institute, New York, New York 10010, USA}
\affiliation{Nano-Bio Spectroscopy Group and ETSF, Universidad del País Vasco UPV/EHU, San Sebastián 20018, Spain}

\date{\today}

\begin{abstract}
Kagome metals host competing electronic orders, including charge-density-wave (CDW) order and superconductivity, shaped by intertwined lattice, electronic-correlation, and kagome-geometric effects. Here, using quantum electrodynamical density functional theory, we identify an equilibrium cavity route for reshaping this balance in the kagome metal CsV$_3$Sb$_5$. An out-of-plane polarized single-mode cavity selectively softens CDW-related phonons, counteracting pressure-induced hardening and extending the CDW instability toward higher pressures. In the high-pressure regime where the CDW instability is otherwise suppressed, cavity coupling redistributes Eliashberg spectral weight toward lower frequencies, enhances the total electron--phonon coupling (EPC), and increases the EPC-based Allen--Dynes estimate of $T_c$. This response originates from a charge-density redistribution induced by the out-of-plane photon mode, which modifies lattice restoring forces and drives the phonon and EPC renormalization. These results establish cavity quantum electrodynamics as a viable equilibrium route for tuning intertwined charge order, lattice dynamics, and superconductivity in kagome materials.
\end{abstract}

\maketitle
The kagome metals AV$_3$Sb$_5$ (A= K, Rb, Cs) have emerged as an important platform for studying intertwined quantum phases arising from kagome electronic structure and collective ordering phenomena.~\cite{Wilson2024NRM_AV3Sb5Review,Neupert2022NatPhys,Hu2023NPJQM,DiSante2025arXiv,Jiang2023NSR_Review,Zhou2026FrontPhys_AV3Sb5TopicalReview}. 
Owing to the presence of Dirac dispersions, van Hove singularities, and nontrivial band topology, these materials exhibit a rich phase diagram involving charge-density-wave (CDW) order, superconductivity, nematicity, and anomalous transport responses.~\cite{Kang2022NatPhys,Zhou2026FrontPhys_AV3Sb5TopicalReview,Hu2022NatCommun,Jiang2023NSR_Review,Wu2021PRL,Nie2022Nature,Guo2022Nature,Le2024Nature}. 
In the AV$_3$Sb$_5$ family, the CDW state strongly reconstructs the low-energy electronic structure and exhibits a close interplay with superconductivity~\cite{Ortiz2020PRL,Liang2021PRX,Li2022NatCommun}.
Pressure studies on CsV$_3$Sb$_5$ further revealed double superconducting domes across the evolution of the CDW phase, suggesting distinct relationships between superconductivity and charge order in different pressure regimes.~\cite{Chen2021PRL,YuNatCommun2021,Roppongi2023NC}
Subsequent NMR/NQR, X-ray scattering, and spectroscopic studies further suggested that this behavior cannot be understood as a simple monotonic suppression of a single CDW order, but instead involves a complex CDW landscape with three-dimensional ordering, interlayer phase degrees of freedom, and possible incommensurate modulations~\cite{Liang2021PRX,zheng2022nature,Li2022NatCommun,Feng2023NPJQM,Stier2024PRL,Wen2023SciBull,shao2026PRL}.
These observations highlight the close interplay between superconductivity and the underlying CDW instability, and raise the central question of how the CDW state can be manipulated to clarify its role in the superconducting phase diagram.

Considerable effort has therefore been devoted to tuning the CDW phase in CsV$_3$Sb$_5$ through hydrostatic pressure, strain engineering, and chemical substitution~\cite{Chen2021PRL,Qian2021PRB_Strain,Frachet2024PRL_Strain,Lin2024NatCommun,liu2022PRBdoping,liu2022PRM}.
These studies demonstrate the exceptional tunability of the CDW order and its strong influence on superconductivity. However, pressure and strain primarily act through global structural modifications, making it difficult to disentangle the intertwined lattice and electronic contributions to the CDW instability. Recently, cavity quantum electrodynamics has emerged as a promising route for engineering collective phenomena in quantum materials through hybrid light-matter coupling without altering crystal parameters~\cite{Lu2025_CavityEngineering,keren2026nature,graziotto2026natphys,Faist2022Science}. Previous model-Hamiltonian studies established important conceptual insights by showing that quantum light can modify electronic correlations and superconductivity, for example through cavity-mediated interactions and reshaped pairing channels~\cite{Li2020PRL,Sentef2018SciAdv,Schlawin2019PRL,wei2025PRL}. However, the microscopic influence of cavity coupling on realistic materials, where electronic structure, lattice dynamics, and electron--phonon coupling are coupled self-consistently, remains largely unexplored.

\begin{figure}
  \centering
  \includegraphics[width=\linewidth]{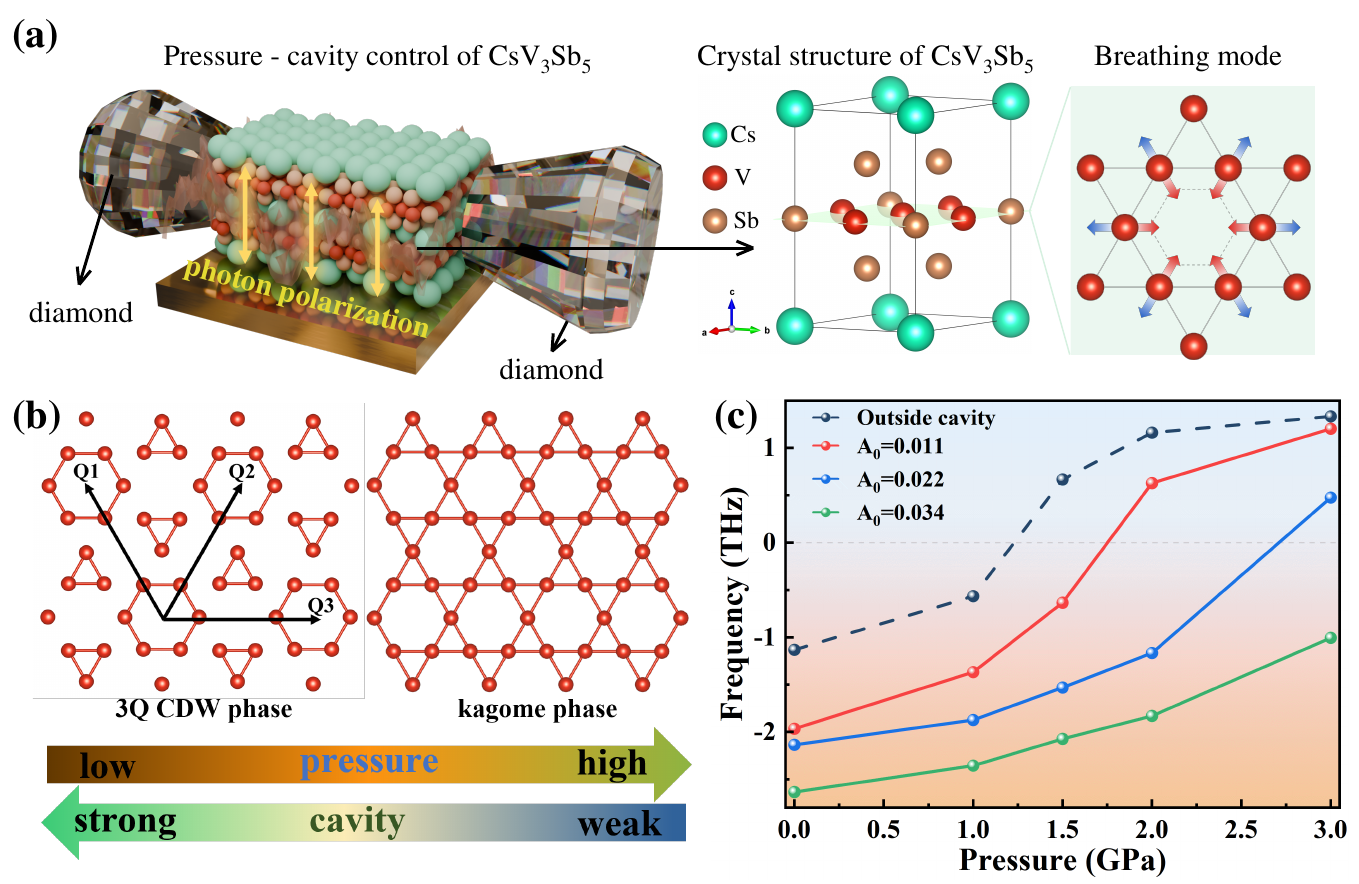}
    \caption{\textbf{Pressure--cavity control of CDW instability in CsV$_3$Sb$_5$} (a)Schematic illustration of the pressure--cavity model for CsV$_3$Sb$_5$. The crystal is placed in a diamond-anvil-cell-like hydrostatic-pressure geometry and coupled to a single confined photon mode polarized along the crystallographic $c$ axis, perpendicular to the kagome layers. The right panels show the crystal structure and the CDW-related V-sublattice breathing vibration.
    (b) Conceptual comparison between the 3$Q$ CDW phase and the normal kagome phase, highlighting the opposite effects induced by pressure and cavity coupling. (c) Pressure dependence of the lowest phonon frequency at the L point for different cavity-field amplitudes $A_0$.}    
    \label{Fig.1}
\end{figure}

Here we address this question in the kagome metal CsV$_3$Sb$_5$ using quantum electrodynamical density functional theory (QEDFT)~\cite{Lu2025_CavityEngineering,PhysRevA.109.052823,Lu2024PNAS_MgB2}, which enables a first-principles description of cavity light--matter coupling in realistic materials. 
We show that an out-of-plane polarized single-mode cavity, modeled as a one-dimensional confined photon field, acts as a equilibrium perturbation to the layered CsV$_3$Sb$_5$ structure and its CDW-related breathing distortion (Fig.~\ref{Fig.1}a). 
Rather than phenomenologically shifting an order parameter, the cavity induces a charge redistribution that renormalizes CDW soft phonons and extends the CDW-related lattice-instability regime toward higher pressures(Fig.~\ref{Fig.1}b,c). 
In the pressure-stabilized regime, the cavity restores the softening of CDW-related phonons, transfers electron--phonon spectral weight to lower frequencies, enhances the total EPC strength, and enhances the EPC-based Allen--Dynes estimate of $T_c$. 
Since the superconducting domes in CsV$_3$Sb$_5$ exhibit distinct relationships with the CDW phase, the ability to manipulate the CDW instability through cavity coupling could potentially provide a route to disentangle the interplay between charge order, lattice dynamics, and superconductivity in kagome materials.
Our results establish cavity engineering as a viable approach for controlling collective ordered phases in quantum materials and open opportunities for investigating intertwined electronic states in cavity-modified kagome systems.

CsV$_3$Sb$_5$ consists of V-kagome layers separated by Cs spacer layers, with Sb atoms embedded in the surrounding V–Sb framework, and the CDW instability is associated with a breathing-like distortion of the V sublattice, involving collective in-plane contraction and expansion of the kagome network [Fig.~\ref{Fig.1}(a)]. 
Hydrostatic pressure and cavity coupling act on this instability in opposite ways: pressure stabilizes the high-symmetry kagome lattice by hardening the CDW soft mode, whereas a z-polarized cavity field, chosen to preserve the in-plane kagome symmetry and avoid explicit cavity-induced symmetry breaking, restores the tendency toward the CDW instability [Fig.~\ref{Fig.1}(b)]. Pressure drives the system from the CDW-distorted phase toward the high-symmetry kagome lattice~\cite{Chen2021PRL,YuNatCommun2021,ZhangPRB2021,SiPRB2022}, whereas the cavity field pushes the lattice back toward the CDW instability.
This antagonism is quantified in Fig.~\ref{Fig.1}(c) through the pressure dependence of the lowest L-point (see Fig.2) phonon frequency for different cavity couplings.
To characterize the strength of the light–matter interaction in a quantitative manner, we parameterize the single-mode cavity coupling by a dimensionless amplitude $A_0$. Within the long-wavelength approximation of our QEDFT implementation, $A_0$ is defined as $A_0=\tilde{\lambda}_{\alpha}/\sqrt{2\tilde{\omega}_{\alpha}}$, where $\tilde{\lambda}_{\alpha}$ and $\tilde{\omega}_{\alpha}$ denote the effective cavity-coupling constant and dressed photon frequency, respectively~\cite{Lu2024PNAS_MgB2,fan2026}. This quantity plays the role of an effective vector-potential amplitude entering the paramagnetic-current coupling. Detailed definitions, numerical settings, and the mapping between $A_0$ and experimental field scales are provided in the Supplemental Material. As shown in Fig.~\ref{Fig.1}(c), increasing $A_0$ systematically delays the pressure-induced removal of the soft-mode instability. We emphasize that this criterion tracks the cavity dependence of the harmonic lattice-instability boundary, rather than the full finite-temperature anharmonic CDW transition line.

\begin{figure}[!b]
  \centering
  \includegraphics[width=\linewidth]{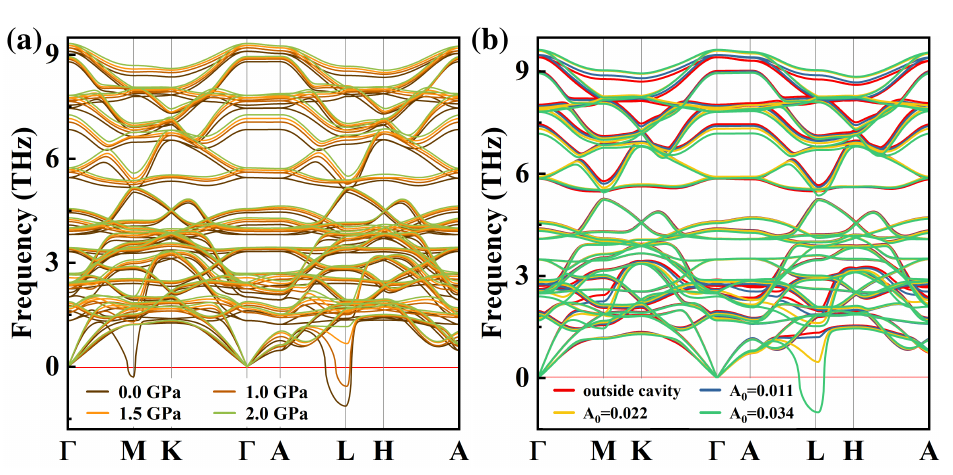}
  \caption{\textbf{Competing effects of pressure and cavity coupling on phonons in CsV$_3$Sb$_5$.}
  (a) Phonon dispersions at different hydrostatic pressures, showing progressive hardening of low-frequency modes and suppression of the CDW instability. (b) Phonon dispersions at 3 GPa under different cavity-field couplings $A_0$, compared with the outside cavity case. The cavity selectively softens low-energy phonons, counteracting the pressure-induced stabilization of the lattice.}
 \label{Fig.2}
\end{figure}

Experimentally, the CDW transition in CsV$_3$Sb$_5$ is rapidly suppressed with pressure and disappears near $\sim$2~GPa, while the superconducting transition temperature \(T_c\) is enhanced and exhibits a dome-like pressure dependence~\cite{Chen2021PRL,YuNatCommun2021}.
Our harmonic phonon calculations capture the pressure-induced suppression of the CDW instability: pressure progressively hardens the CDW soft branches, with the M-point instability disappearing first and a residual L-point instability persisting up to 1 GPa [Fig.~\ref{Fig.2}(a)], in qualitative agreement with experiment and consistent with prior first-principles studies~\cite{ZhangPRB2021}. Above 1.5 GPa, all imaginary phonon frequencies disappear, indicating that the high-symmetry phase becomes dynamically stable within harmonic lattice dynamics.
In contrast, embedding CsV$_3$Sb$_5$ in a single-mode cavity with $E \parallel \hat{z}$ produces the opposite effect to pressure-induced stabilization. As shown in Fig. 2(b), increasing the light–matter coupling strength systematically softens the CDW-relevant L-point branch at 3 GPa, where the free-space structure is already dynamically stable. The special sensitivity of the L-point branch is notable because the L instability carries both the in-plane V-breathing distortion and the out-of-plane stacking phase degree of freedom.
At sufficiently large coupling, the L-point mode is driven back into an imaginary-frequency instability, whereas higher-energy phonon branches remain essentially unchanged.
This behavior demonstrates the mode selectivity of the cavity-induced phonon renormalization: pressure hardens the CDW soft mode, whereas the cavity field selectively softens the same mode and delays its pressure-induced suppression.
\begin{figure}
  \centering
  \includegraphics[width=\linewidth]{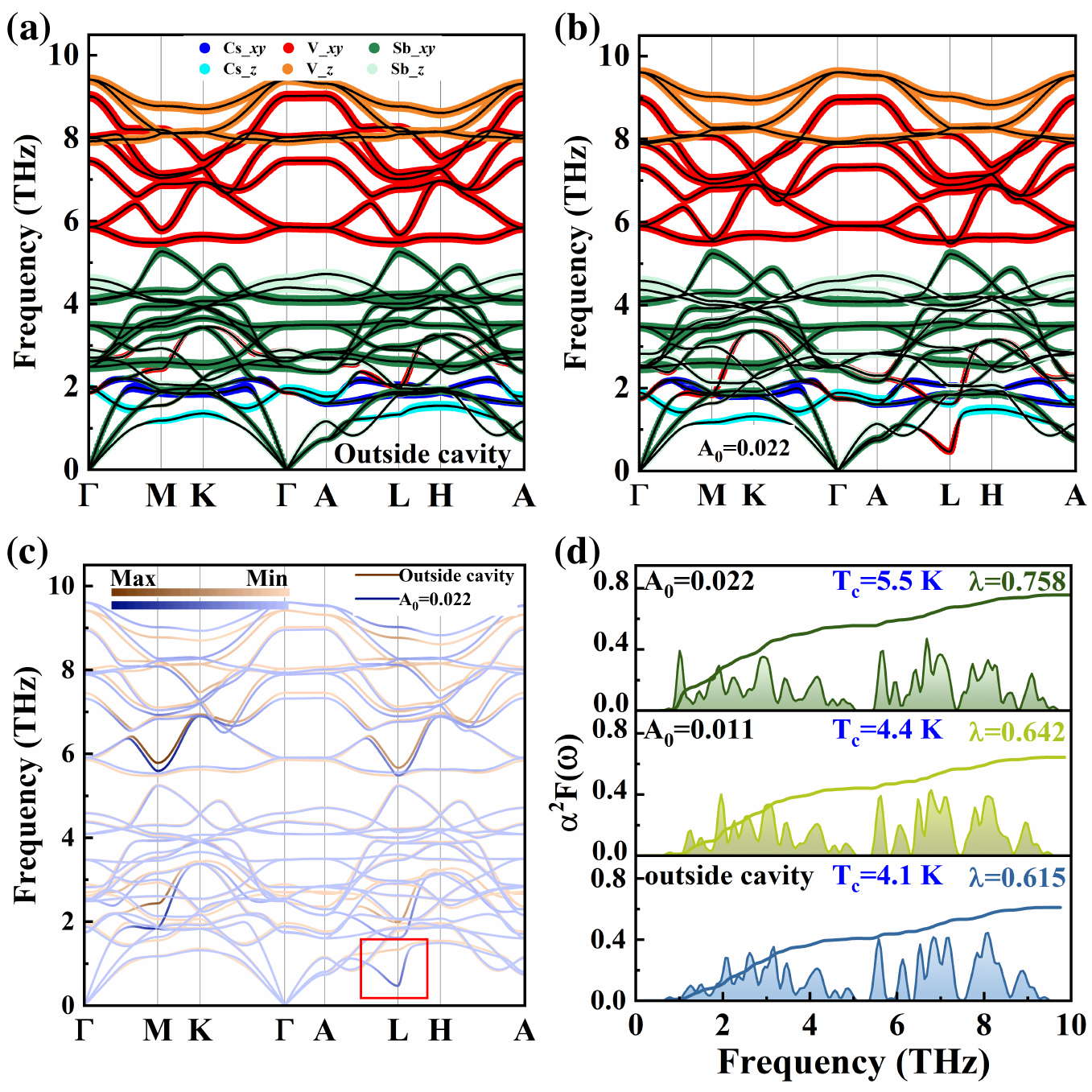}
  \caption{\textbf{Cavity-enhanced electron--phonon coupling and superconductivity in CsV$_3$Sb$_5$.}
  (a) and (b) Phonon dispersions at 3~GPa for free space and cavity couplings, with mode characters projected onto in-plane ($xy$) and out-of-plane ($z$) vibrations of Cs, V, and Sb.
  (c) Phonon dispersions with linewidths mapped onto the branches, highlighting momentum-selective enhancement of electron--phonon scattering.
  (d) Eliashberg spectral function $\alpha^2F(\omega)$ and the cumulative EPC $\lambda(\omega)$.}
  \label{Fig.3}
\end{figure}

The CDW instability in CsV$_3$Sb$_5$ is closely associated with V-atom breathing vibrations in the kagome lattice~\cite{luo2022,ratcliff2021,ortiz2021PRX}. 
To characterize the cavity-induced renormalization microscopically, Figs.~\ref{Fig.3}(a,b) compare the phonon dispersions of CsV$_3$Sb$_5$ at 3 GPa outside the cavity and under a representative cavity coupling. The phonon modes are projected onto atomic species and decomposed into in-plane and out-of-plane vibrational components, allowing the character of the softened branch to be identified.
While the free-space phonon spectrum at 3 GPa is dynamically stable [Fig.~\ref{Fig.3}(a)], increasing the cavity coupling progressively softens a zone-boundary mode at the L point. At $A_0 = 0.022$, the mode is substantially softened but remains real, providing a useful case for analyzing the cavity-induced change in phonon character before the onset of an imaginary-frequency instability [Fig.~\ref{Fig.3}(b)]. 
The mode projections reveal a cavity-induced change of phonon character near the L point. In free space, the lowest-energy branch is primarily Cs-dominated [the cyan curves in Fig.~\ref{Fig.3}(a)], whereas under the cavity coupling the softened branch becomes strongly V-dominated with in-plane character [the red curves in Fig.~\ref{Fig.3}(a)], corresponding to the CDW-related breathing vibration illustrated schematically in Fig.~\ref{Fig.1}(a). This reflects hybridization and an avoided crossing between Cs- and V-derived phonon branches, leading to a transfer of mode character at the zone boundary. Through this mechanism, the cavity effectively pulls the CDW-related V breathing mode down in energy, thereby promoting the lattice instability. Importantly, the cavity does not create a new lattice instability, but selectively re-softens the preexisting CDW-related breathing mode that survives as a low-energy fluctuation in the pressure-stabilized phase. The cavity therefore reshapes an existing CDW instability landscape rather than imposing an unrelated structural distortion. Consistent with the expectation that cavity QED can renormalize effective EPC and thereby influence superconductivity in equilibrium~\cite{Sentef2018SciAdv,Lu2024PNAS_MgB2}, the cavity selectively enhances the low-frequency electron–phonon response by further softening CDW-proximate modes near the Brillouin-zone boundary, while leaving the high-energy phonon spectrum largely unchanged.

Momentum-dependent EPC is a key ingredient in the CDW physics of $A$V$_3$Sb$_5$~\cite{XiePRB2022,LuoNatCommun2022,UykurNPJQM2022,Zhong2023NC}. The cavity-induced modification of the phonon spectrum is further clarified in Fig.~\ref{Fig.3}(c), where the color scale represents the mode-resolved phonon linewidth, reflecting the electron–phonon scattering strength associated with each phonon mode.
In free space, the linewidth is predominantly concentrated near the Brillouin-zone boundary, with pronounced contributions around the M and L points at frequencies of a few THz, indicating that EPC is already strongest in these CDW-relevant regions. 
Under cavity coupling, the linewidth distribution is modified in a highly localized manner: enhanced linewidth appears on the softened low-frequency branch near the L point, while most other modes remain largely unchanged.
Combined with the mode projections in Figs.~\ref{Fig.3}(a,b), this indicates that the enhanced low-energy EPC is associated with a mode that evolves from a predominantly Cs-derived branch in free space into a V-dominated in-plane vibration under cavity coupling. 
Thus, the cavity not only softens the phonon spectrum but also redistributes EPC toward CDW-relevant V-site dynamics at the zone boundary.

The impact on EPC and the estimated superconductivity is evident from Fig.~\ref{Fig.3}(d), and the EPC and Allen--Dynes~\cite{AllenDynesPRB1975} $T_c$  reported here are evaluated for the pressure-stabilized high-symmetry primitive cell at 3 GPa.The cavity-induced softening redistributes the Eliashberg spectral function $\alpha^2F(\omega)$ toward low frequencies, enhancing the spectral weight associated with CDW-proximate phonons while leaving the high-energy part of the spectrum largely unaffected.
This redistribution results in a systematic enhancement of the total electron-phonon strength $\lambda$, from 0.615 outside the cavity to 0.642 at $A_0 = 0.011$ and 0.758 at $A_0 = 0.022$.
Within the Allen--Dynes estimate, and for the pressure-stabilized high-symmetry phase where the CDW instability is fully suppressed, this enhanced low-frequency coupling increases the superconducting transition temperature from $T_c = 4.1$ K to 4.4 K and 5.5 K, respectively. This trend is broadly consistent with recent ARPES-based Eliashberg analyses showing that CsV$_3$Sb$_5$ hosts an intermediate EPC strength and that strengthening V-derived EPC channels can be associated with an enhanced superconducting transition temperature~\cite{Zhong2023NC}.
These results indicate that the cavity enhances the EPC-driven superconducting tendency primarily through mode-selective phonon softening and the associated increase in low-energy EPC, consistent with minimal modification of the electronic structure, as evidenced by the nearly unchanged density of states and band structure (shown in Fig. S1 and Fig. S2).
The full $A_0$ dependence of the superconducting parameters is shown in Fig. S3, where the EPC constant $\lambda$, logarithmic average frequency $\omega_{\log}$, and Allen–Dynes $T_c$ are plotted as functions of cavity coupling. With increasing $A_0$, $\lambda$ increases monotonically, whereas $\omega_{\log}$ decreases because the Eliashberg spectral weight shifts toward lower frequencies. Nevertheless, $T_c$ increases, showing that the enhancement of $\lambda$ outweighs the reduction of $\omega_{\log}$. This confirms that the cavity-enhanced superconductivity is driven by strengthened low-energy electron–phonon coupling rather than by an overall stiffening of the phonon spectrum.

\begin{figure}
  \centering
  \includegraphics[width=\linewidth]{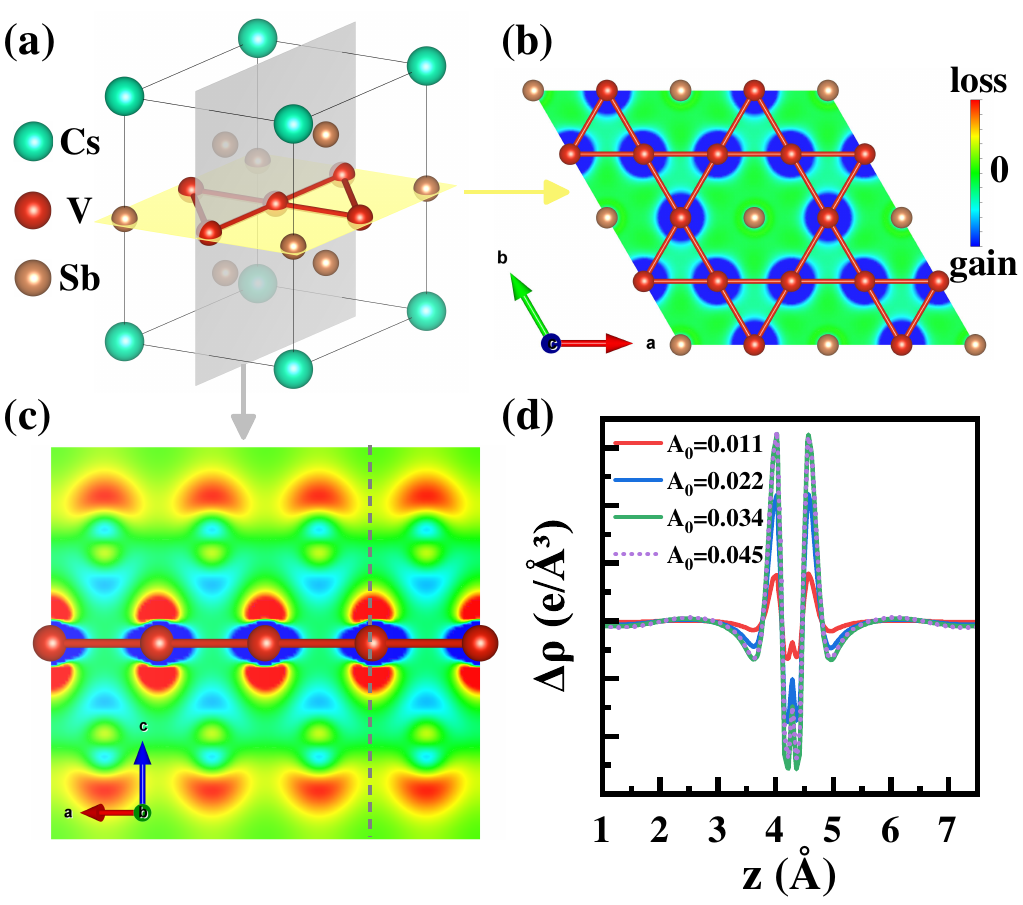}
  \caption{\textbf{Cavity-induced charge redistribution in CsV$_3$Sb$_5$.} 
  (a) Crystal-structure schematic of CsV$_3$Sb$_5$, highlighting the V kagome layer and the out-of-plane plane used for the charge-density analysis.
  (b) Two-dimensional map of $\Delta\rho(\mathbf{r})$ in the $ab$ plane across the V kagome layer. 
  (c) Two-dimensional map of $\Delta\rho(\mathbf{r})$ in the $bc$ plane, highlighting the out-of-plane component of the cavity-induced charge response. The gray dashed line indicates the line-cut path used for the profiles in (d).
  (d) Line profiles of $\Delta\rho(\mathbf{r})$ extracted along the dashed line in (c), i.e., along the out-of-plane ($z$) direction through the V atom, for different cavity couplings \(A_0\).}  
    \label{Fig.4}
\end{figure}

The isosurfaces [Fig.~\ref{Fig.4}] reveal a pronounced dipolar rearrangement aligned with the cavity polarization, with charge accumulating and depleting in alternating lobes above and below the kagome layer, predominantly around the V–Sb network, via dynamical localization~\cite{Lu2024PNAS_MgB2,dongbin2025}. This pattern reflects an out-of-plane electronic polarization that may couple to the three-dimensional CDW stacking in CsV$_3$Sb$_5$~\cite{TanPRL2021,LiangPRX2021,JinPRL2024}. 
Real-space maps in the kagome plane and vertical cross sections [Figs.~\ref{Fig.4}(b) and \ref{Fig.4}(c)] further show that the redistribution is spatially modulated on the scale of the V–Sb coordination environment, indicating a local rearrangement of charge rather than a spatially uniform redistribution.

%
In particular, besides the enhanced electron accumulation in the V–V plane, the charge response is also markedly asymmetric around the Sb sites along the out-of-plane direction. Because Sb atoms form the immediate coordination environment of the V kagome network and participate in the three-dimensional CDW stacking pattern, this Sb-centered asymmetry suggests that the cavity modifies the local electrostatic environment of the V–Sb framework in a strongly anisotropic manner. In this sense, the cavity field produces a local, polarization-selective perturbation that is distinct from hydrostatic compression.
This redistribution strengthens local electronic screening in the V kagome plane and is consistent with a reduction of the effective restoring forces associated with V–V and V–Sb coordinated distortions underlying the CDW-proximate zone-boundary modes. 
Such a reduction offers a natural real-space mechanism for the selective softening and enhancement of the preexisting CDW-related V breathing mode, most prominently the cavity-enhanced L-point soft mode associated with the distortion pattern illustrated in Fig.~\ref{Fig.1}(a) (Figs.~\ref{Fig.3}).
The line cuts in Fig.~\ref{Fig.4}(d) confirm that $\Delta\rho$ grows systematically with $A_0$, establishing a direct and tunable electronic response. 
This out-of-plane charge rearrangement provides a consistent microscopic picture for the selective renormalization and reinforcement of the preexisting CDW-related breathing instabilities at M and L (Fig.~\ref{Fig.2}), and in the pressure-stabilized regime, for the enhancement of EPC in the same low-frequency window that governs $\alpha^2F(\omega)$ and $T_c$ (Fig.~\ref{Fig.3}).

We demonstrate that a optical cavity provides a symmetry-selective equilibrium control knob for competing lattice and electronic instabilities in the kagome metal CsV$_3$Sb$_5$, distinct from conventional pressure tuning.
The central result is a pressure--cavity antagonism of the CDW-related soft branch: hydrostatic compression hardens the CDW soft mode and removes imaginary frequencies, whereas cavity coupling softens the same instability and shifts its suppression to higher pressures.
In the high-pressure regime where the CDW instability is otherwise suppressed, cavity coupling enhances the EPC-based superconducting tendency by increasing the low-frequency weight of $\alpha^2F(\omega)$ and the total EPC constant $\lambda$, thereby raising the Allen--Dynes estimate of $T_c$.
Microscopically, this behavior is associated with a cavity-induced out-of-plane charge redistribution centered on the V--Sb network, which modifies the effective lattice restoring forces of the CDW-related zone-boundary phonon modes and reshapes the EPC contribution near the CDW wave vectors.
By providing a means to modify the CDW instability without relying solely on global structural compression, our results offer a new route to interrogate the unresolved relationship between CDW order and superconductivity in CsV$_3$Sb$_5$.

\begin{acknowledgments}
We acknowledges the support by the National Key Research and Development Program of China (2022YFA1403500), the National Natural Science Foundation of China (Grant No. 12474174), Hangzhou Tsientang Education Foundation and the Max Planck Partner group programme.
D.S. was supported by the National Research Foundation of Korea (NRF) grant funded by the Korea government (MSIT) (No. RS-2024-00333664) and the Ministry of Science and ICT(No. RS-2022-NR068223). This work was supported by the European Research Council (ERC Synergy Grant Agreement No. 101167294 (UnMySt)), the Cluster of Excellence “CUI: Advanced Imaging of Matter” of the Deutsche Forschungsgemeinschaft (DFG)—EXC 2056—Project ID 390715994 and Grupos Consolidados y Alto Rendimiento UPV/EHU, Gobierno Vasco (IT1453-22). The Flatiron Institute is a division of the Simons Foundation.
The authors declare no competing interests.
\end{acknowledgments}
\nocite{*}

\bibliography{apssamp}

\end{document}


\title{Supplemental Material for:\\
\textit{“Cavity Tuning of the CDW--Superconductivity Interplay in a Kagome Metal”}}

\author{Lan-Ting Shi}
\thanks{These authors contributed equally to this work.}
\affiliation{Tsientang Institute for advaced study, Zhejiang 310024, China}
\affiliation{Songshan Lake Materials Laboratory, Guangdong 523808, China}

\author{I-Te Lu}
\thanks{These authors contributed equally to this work.}
\affiliation{Max Planck Institute for the Structure and Dynamics of Matter and Center for Free-Electron Laser Science, Luruper Chaussee 149, Hamburg 22761, Germany}

\author{Xin Wang}
\affiliation{Songshan Lake Materials Laboratory, Guangdong 523808, China}
\affiliation{Department of Materials Science and Engineering,
City University of Hong Kong, Kowloon 999077, Hong Kong}

\author{Dongbin Shin}
\email{dshin@gist.ac.kr}
\affiliation{Department of Physics and Photon Science, Gwangju Institute of Science and Technology (GIST), Gwangju 61005, Republic of Korea}
\affiliation{Max Planck Institute for the Structure and Dynamics of Matter and Center for Free-Electron Laser Science, Luruper Chaussee 149, Hamburg 22761, Germany}

\author{Lede Xian}
\email{lede.xian@mpsd.mpg.de}
\affiliation{Tsientang Institute for advaced study, Zhejiang 310024, China}
\affiliation{Songshan Lake Materials Laboratory, Guangdong 523808, China}
\affiliation{Max Planck Institute for the Structure and Dynamics of Matter and Center for Free-Electron Laser Science, Luruper Chaussee 149, Hamburg 22761, Germany}

\author{Angel Rubio}
\affiliation{Max Planck Institute for the Structure and Dynamics of Matter and Center for Free-Electron Laser Science, Luruper Chaussee 149, Hamburg 22761, Germany}

\date{\today}

\date{\today}

\maketitle

\subsection{Computational details}

First-principles calculations for CsV$_3$Sb$_5$ outside the cavity were performed within density-functional theory (DFT) using the Quantum ESPRESSO package~\cite{Giannozzi_2009}. We employed the Perdew--Burke--Ernzerhof (PBE) generalized--gradient approximation~\cite{PhysRevLett.77.3865} together with ultrasoft pseudopotentials~\cite{PhysRevB.41.7892}, and included van der Waals interactions through the DFT-D3 correction \cite{10.1063/1.3382344} to capture the weak interlayer coupling of this layered kagome material. The kinetic-energy cutoffs for the wave functions and charge density were set to 60 Ry and 600 Ry, respectively. 
Structural relaxations were carried out at each hydrostatic pressure until the residual forces, total-energy changes, and residual stress were converged to below $1.0 \times 10^{-5}$ Ry/Bohr, $1.0 \times 10^{-6}$ Ry, and $0.5$ kbar, respectively, corresponding to the force, energy, and pressure convergence thresholds used in the \texttt{vc-relax} calculations. The optimized lattice parameters of the outside cavity structures from 0 to 3 GPa are summarized in Table~S1.
Brillouin-zone integrations used Monkhorst--Pack k-point meshes of 10$\times$10$\times$6 for structural relaxations and self-consistent calculations. Metallic occupations were treated using Gaussian smearing with a smearing width of 0.01 Ry. 

Phonon properties and electron--phonon coupling (EPC) calculations were computed within density-functional perturbation theory (DFPT)~\cite{RevModPhys.73.515}, using a 20$\times$20$\times$12 k-point grid and a 5$\times$5$\times$3 q-point grid for the EPC calculations. Dynamical matrices were evaluated on a uniform $q$-point grid and interpolated along the high-symmetry path $\Gamma$--M--K--$\Gamma$--A--L--H--A. A charge-density-wave (CDW) instability was identified by the appearance of imaginary phonon frequencies, which are reported as negative values, with particular emphasis on the soft branches near the M and L points that are relevant to the 2$\times$2$\times$2 lattice distortions in CsV$_3$Sb$5$. From the DFPT electron–phonon matrix elements, we constructed the Eliashberg spectral function $\alpha^2F(\omega)$ and the EPC constant $\lambda$, and computed the logarithmic average frequency $\omega_{\log}$. The superconducting critical temperature $T_c$ was estimated using the Allen–Dynes modified McMillan formula~\cite{PhysRevB.12.905},
\begin{equation}
T_{\mathrm{c}} = \frac{\omega_{\mathrm{log}}}{1.2} \exp\left[-\frac{1.04(1 + \lambda)}{\lambda - \mu^{*} - 0.62\lambda\mu^{*}}\right],
\label{eq:allen_dynes}
\end{equation}
where $\mu{*}=0.10$ is the parameter for the screened Coulomb pseudopotential. Unless otherwise stated, we use the conventional value $\mu{*}=0.10$ for estimating $T_c$~\cite{PhysRevB.12.905}. The cumulative EPC constant is evaluated as
\begin{equation}
\lambda(\omega) = 2 \int_{0}^{\omega} \frac{\alpha^{2}F(\omega')}{\omega'} d\omega'.
\label{eq:lambda_integral}
\end{equation}
The EPC constant \( \lambda \) used in Eq.~\eqref{eq:allen_dynes} is \( \lambda(\omega_{\mathrm{max}}) \), where \( \omega_{\mathrm{max}} \) is the maximum of the phonon frequency.The logarithmic average phonon frequency is defined as
\begin{equation}
\omega_{\mathrm{log}} = \exp\left[\frac{2}{\lambda} \int \frac{d\omega}{\omega} \alpha^{2}F(\omega) \ln \omega\right].
\label{eq:omega_log}
\end{equation}
The Eliashberg spectral function \( \alpha^{2}F(\omega) \) is calculated as
\begin{equation}
\alpha^{2}F(\omega) = \frac{1}{2\pi N(E_{\mathrm{F}})} \sum_{q\nu} \delta(\omega - \omega_{q\nu}) \frac{\gamma_{q\nu}}{\hbar \omega_{q\nu}},
\label{eq:eliashberg}
\end{equation}
where \( \omega_{q\nu} \) are phonon frequencies, and \( \gamma_{q\nu} \) is the phonon linewidth described by
\begin{equation}
\gamma_{q\nu} = 2\pi \omega_{q\nu} \sum_{ij} \int \frac{d^{3}k}{\Omega_{\mathrm{BZ}}} |g_{q\nu}(k,i,j)|^{2}
\times \delta(\epsilon_{q,i} - \epsilon_{\mathrm{F}}) \delta(\epsilon_{k+q,j} - \epsilon_{\mathrm{F}}).
\label{eq:phonon_linewidth}
\end{equation}
Here, \( g_{q\nu}(k, i, j) \) is the electron--phonon coupling matrix element, and \( \epsilon_{q,i} \) is the electronic energy at wavevector \( q \) and band index \( i \).

Cavity-coupled calculations were performed using our in-house implementation of \emph{ab initio} quantum-electrodynamical density-functional theory (QEDFT), based on the electron–photon exchange approximation within the local-density approximation, which is derived within the Pauli–Fierz framework in the long-wavelength approximation and the velocity gauge. In this framework, cavity vacuum fluctuations enter through fluctuations of the electronic paramagnetic current, giving rise to an additional electron–photon exchange–correlation potential, $v_{\mathrm{pxc}}(\mathbf{r})$, in the Kohn–Sham equations. We consider a single-mode cavity with out-of-plane polarization, $\mathbf{E}\parallel\hat{z}$, and parameterize the light–matter coupling by the dimensionless cavity amplitude $A_0$ used throughout the main text. Importantly, the cavity contribution was included consistently in both the ground-state self-consistent calculations performed with QE \texttt{pw.x} and the linear-response DFPT calculations performed with QE \texttt{ph.x}, including the density response of $v_{\rm pxc}$. Except for this additional cavity-induced term and its response, all computational settings, including the exchange–correlation functional, pseudopotentials, dispersion correction, plane-wave cutoffs, and k/q sampling, were kept identical to the corresponding calculations outside the cavity, enabling a controlled comparison between cavity-free and cavity-coupled results.

To describe the coupling between quantized photon fields and the electronic structure from first principles, we employ the QEDFT framework based on the nonrelativistic Pauli–Fierz (PF) Hamiltonian in the Coulomb gauge and long-wavelength limit~\cite{PhysRevA.109.052823}. For a set of $M_p$ effective photon modes, the minimal-coupling PF Hamiltonian in terms of dressed photon modes can be written as
\begin{equation}
\hat{H}_{\mathrm{PF}} = \hat{H}_{\mathrm{M}} + \frac{1}{c}\hat{\mathbf{J}}_p \cdot \hat{\mathbf{A}} 
+ \sum_{\alpha=1}^{M_p} \hbar \tilde{\omega}_{\alpha} \left( \hat{a}_{\alpha}^{\dagger}\hat{a}_{\alpha} + \frac{1}{2} \right),
\label{eq:pf_hamiltonian}
\end{equation}
where $\hat{H}_{\mathrm{M}}$ is the matter Hamiltonian for electrons and nuclei, and $\hat{a}_{\alpha}^{\dagger}$ ($\hat{a}_{\alpha}$) creates (annihilates) a dressed photon in mode $\alpha$. 
The electronic paramagnetic current operator is
\begin{equation}
\hat{\mathbf{J}}_p = \sum_{i=1}^{N_e} (-i\hbar \nabla_i),
\label{eq:current_operator}
\end{equation}
and the vector-potential operator is given by
\begin{equation}
\hat{\mathbf{A}} = c\sum_{\alpha=1}^{M_p} \frac{\tilde{\lambda}_\alpha}{\sqrt{2\tilde{\omega}_\alpha}}
\left(\hat{a}_{\alpha}^{\dagger}+\hat{a}_{\alpha}\right)\boldsymbol{\tilde{\epsilon}}_\alpha,
\label{eq:vector_potential}
\end{equation}
where $\tilde{\lambda}_\alpha$, $\tilde{\omega}_\alpha$, and $\boldsymbol{\tilde{\epsilon}}_\alpha$ denoting the effective light--matter coupling, photon frequency, and polarization of the dressed mode $\alpha$, respectively. The connection between dressed and bare photon parameters is discussed in Refs.~\cite{PhysRevA.109.052823,pnas.2415061121}.

Solving the Pauli–Fierz Hamiltonian in Eq.~\eqref{eq:pf_hamiltonian} directly is computationally demanding. Within QEDFT, the ground-state properties are obtained within an auxiliary Kohn–Sham (KS) scheme, where the KS Hamiltonian is
\begin{equation}
\hat{H}_{\mathrm{KS}} = -\frac{\hbar^2}{2m_e}\nabla^2 + v_{\mathrm{KS}}(\mathbf{r}),
\end{equation}
and the KS potential is decomposed as
\begin{equation}
v_{\mathrm{KS}}(\mathbf{r}) = v_{\mathrm{ext}}(\mathbf{r}) + v_{\mathrm{Hxc}}(\mathbf{r}) + v_{\mathrm{pxc}}(\mathbf{r}).
\label{eq:pxLDA}
\end{equation}
Here $v_{\mathrm{ext}}$ is the external potential, $v_{\mathrm{Hxc}}$ is the usual Hartree plus electronic exchange--correlation potential, and $v_{\mathrm{pxc}}$ is the electron--photon exchange--correlation potential. 
For the electron–photon exchange–correlation potential $v_{\mathrm{pxc}}$, we adopt the pxLDA functional, i.e., the local-density approximation for the electron–photon exchange term; this approximation works well for high bare photon frequency or large light-matter couplings~\cite{PhysRevA.109.052823}. Within this framework, $v_{\mathrm{pxLDA}}(\mathbf{r})$ is obtained by solving
\begin{equation}
\nabla^2 v_{\text{pxLDA}}(\mathbf{r}) = 
-\sum_{\alpha=1}^{M_p} \frac{2\pi^2\tilde{\lambda}_{\alpha}^2}{\tilde{\omega}_{\alpha}^2} 
\left[ (\boldsymbol{\tilde{\epsilon}}_\alpha \cdot \nabla)^2 
\left( \frac{3\rho(\mathbf{r})}{8\pi} \right)^{\frac{2}{3}} \right],
\label{eq:pxlda_poisson}
\end{equation}
where $\rho(\mathbf{r})$ is the electronic density. This term introduces a polarization-dependent correction to the effective potential, thereby coupling the electronic density gradients to the cavity field.

In the pxLDA approximation, for a fixed polarization the cavity-induced correction depends on the effective coupling parameters through combinations such as \(\tilde{\lambda}_{\alpha}^2/\tilde{\omega}_{\alpha}^2\), as shown in Eq.~\eqref{eq:pxLDA}.  
Throughout the main text, we parameterize the light--matter coupling by the dimensionless vector-potential amplitude
\begin{equation}
A_0 \equiv \frac{\tilde{\lambda}_\alpha}{\sqrt{2\tilde{\omega}_\alpha}},
\end{equation}
which corresponds to the mode-amplitude prefactor of the vector-potential operator in Eq.~\eqref{eq:vector_potential}. Ionic forces are evaluated using the Hellmann--Feynman theorem within the same QEDFT framework.

\section{Additional Electronic-Structure Results}
Figure~S1 shows the projected density of states (PDOS) of CsV$_3$Sb$_5$ in the low-energy window from $-2$ to $2$~eV relative to the Fermi level. Figure~S1(a) compares the PDOS outside the cavity under hydrostatic pressures from 0 to 2.0~GPa. The overall low-energy electronic structure is largely preserved upon compression, with only minor changes in the peak intensities and positions. The states near the Fermi level are dominated by V-$d$ orbitals, while the Sb contribution remains relatively weak in this energy range. No pronounced gap opening or abrupt reconstruction of the DOS is observed at $E_F$.Figure~S1(b) shows the PDOS at 3~GPa for different cavity coupling strengths $A_0$. Similar to the pressure-dependent case, the PDOS profiles remain nearly unchanged as $A_0$ increases. In particular, the V-$d$ dominated states around $E_F$ exhibit only weak variations, indicating that the applied $z$-polarized cavity does not substantially reconstruct the low-energy electronic structure. Therefore, the cavity-induced changes in the phonon spectrum and electron--phonon coupling discussed in the main text are unlikely to originate from a large modification of the DOS near $E_F$, but rather from the selective renormalization of lattice instabilities and low-energy phonon modes.

\begin{figure}[H]
  \centering
  \includegraphics[width=\linewidth]{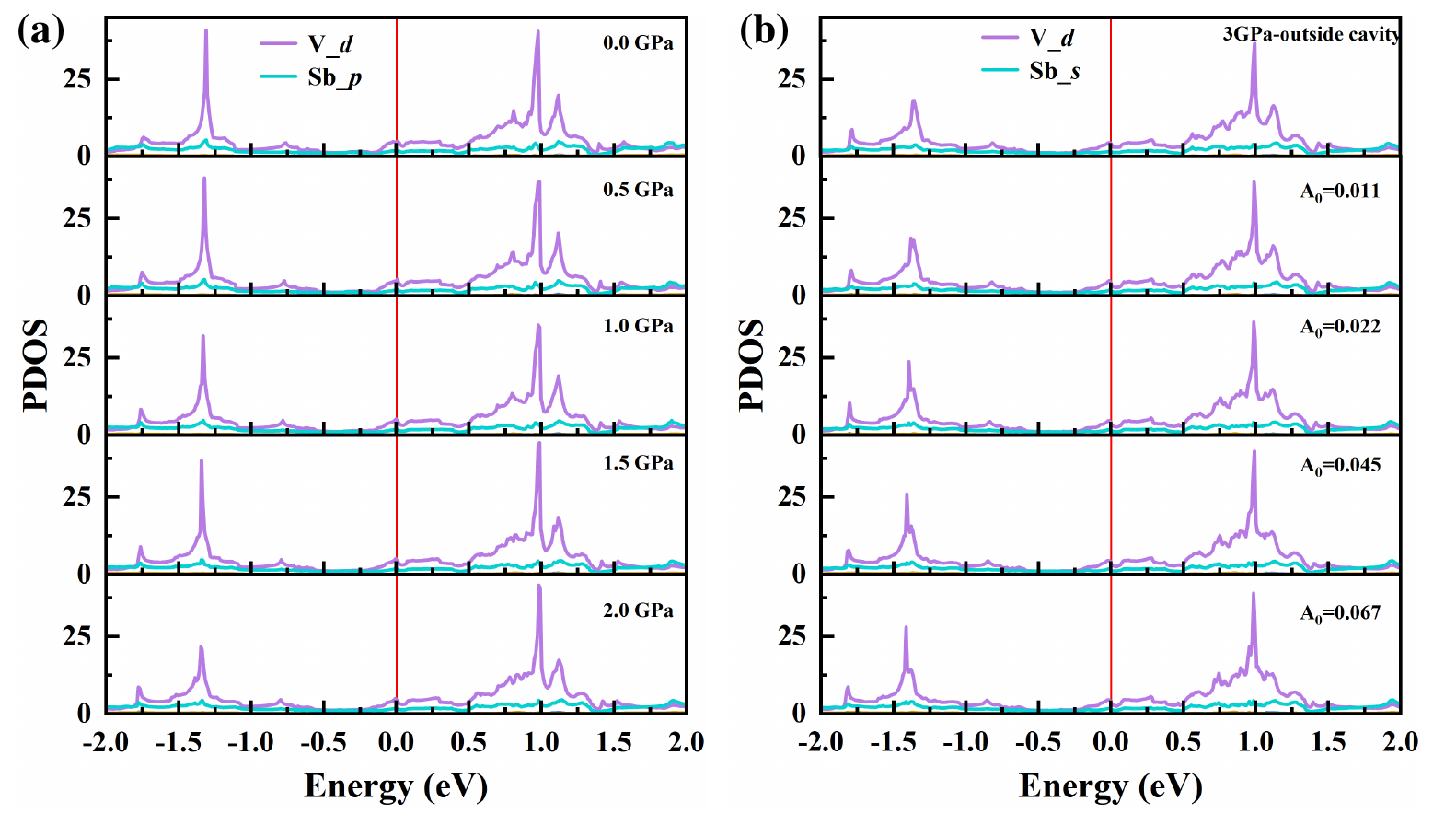}
  \caption{Electronic density of states (DOS) of CsV$_3$Sb$_5$ under different conditions. (a) Pressure dependence of the DOS at 0, 0.5, 1.0, 1.5, and 2.0~GPa in the absence of a cavity. (b) Cavity strength dependence of the DOS at 3~GPa for different light--matter coupling parameters $A_0$. The evolution illustrates how external pressure and cavity modulation independently tune the electronic structure near the Fermi level.}  
    \label{Fig.S1}
\end{figure}

Figure~S2 shows the electronic band structures of CsV$_3$Sb$_5$ at 3~GPa for different cavity coupling strengths $A_0$ along the high-symmetry path $\Gamma$--M--K--$\Gamma$--A--L--H--A. The band dispersions within the plotted energy window from $-1$ to $1$~eV are nearly unchanged upon increasing $A_0$. In particular, the bands crossing the Fermi level remain almost overlapping for the cavity-free and finite-coupling cases, and no obvious gap opening or pronounced band flattening can be identified from the calculated band structures.This result indicates that the applied $z$-polarized cavity does not substantially reconstruct the low-energy electronic band structure near $E_F$. Therefore, the cavity-induced re-softening of the CDW-related phonon modes and the enhancement of the low-frequency electron--phonon coupling discussed in the main text are unlikely to originate from a large modification of the electronic dispersion. Instead, the effect is more consistently understood as a selective renormalization of lattice instabilities and phonon-mediated coupling under cavity light--matter interaction.

\begin{figure}[H]
\centering
\includegraphics[scale=0.5]{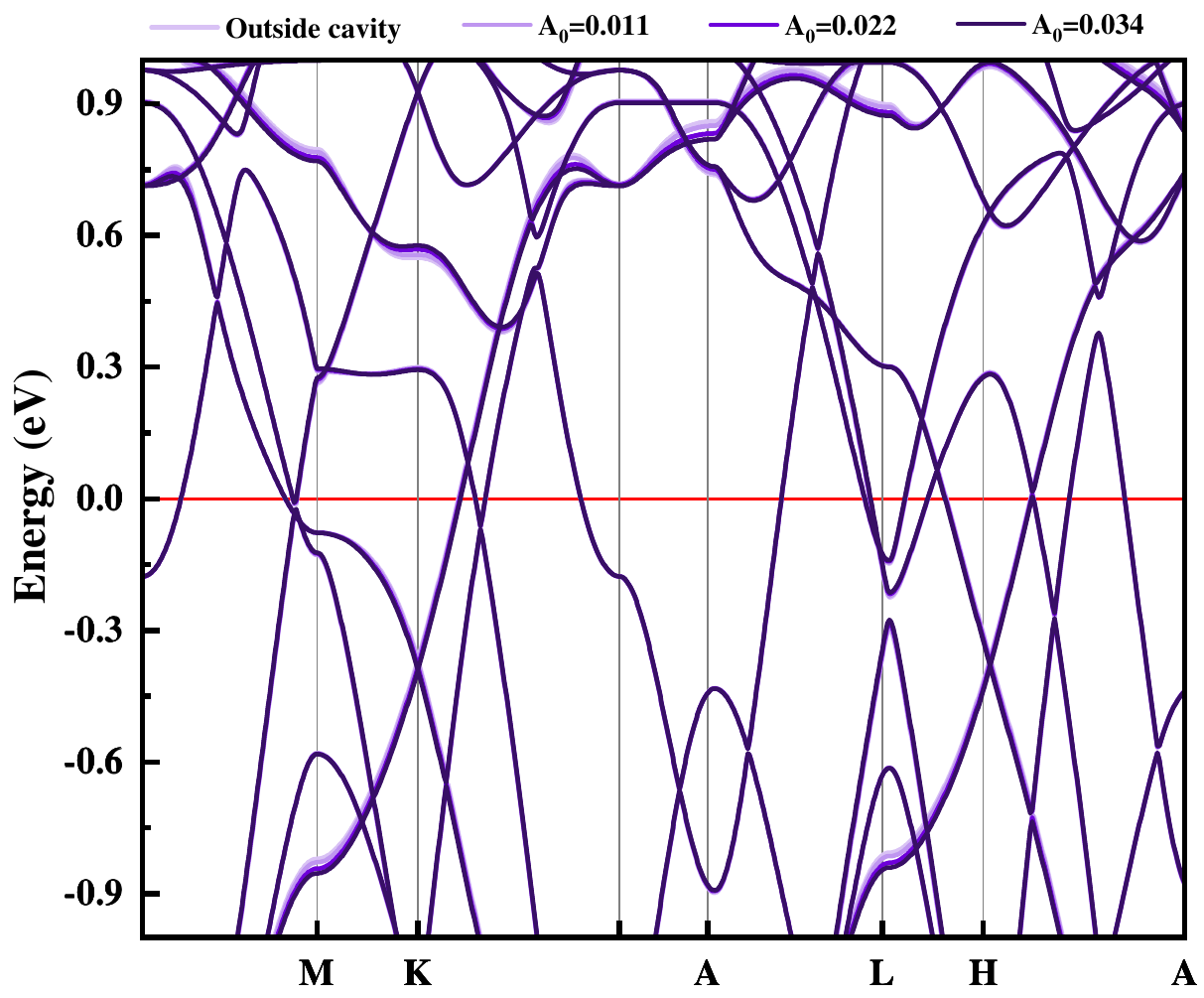}
\caption{Figure~S2. Electronic band structures of CsV$_3$Sb$_5$ at 3~GPa under different cavity coupling strengths $A_0$. The Fermi level set to zero. The band dispersions remain nearly unchanged with increasing $A_0$, indicating that the $z$-polarized cavity does not substantially reconstruct the low-energy electronic structure near $E_F$.}
\label{Fig.S2}
\end{figure}

\section{Evolution of EPC parameters under cavity coupling}
To further quantify the effect of cavity coupling on superconductivity, we analyze the evolution of the electron--phonon coupling constant $\lambda$, the logarithmic average phonon frequency $\omega_{\log}$, and the superconducting transition temperature $T_c$ as a function of the cavity-field amplitude $A_0$, as shown in Fig.~S3. With increasing $A_0$, the total EPC constant $\lambda$ increases monotonically. This behavior originates from the cavity-induced softening of low-energy phonon modes, which enhances the electron--phonon interaction strength in the low-frequency region. At the same time, the logarithmic average frequency $\omega_{\log}$ decreases with increasing $A_0$, reflecting the redistribution of the Eliashberg spectral function $\alpha^2F(\omega)$ toward lower frequencies. This trend is consistent with the phonon softening discussed in the main text. Despite the reduction in $\omega_{\log}$, the superconducting transition temperature $T_c$, estimated using the Allen--Dynes formula, increases with $A_0$. 
This indicates that the enhancement of $\lambda$ dominates over the decrease in $\omega_{\log}$ in determining $T_c$. These results confirm that the cavity-induced enhancement of superconductivity is primarily driven by the increase in low-frequency electron--phonon coupling, rather than by changes in the overall phonon energy scale.
\begin{figure}
  \centering
\includegraphics[scale=0.5]{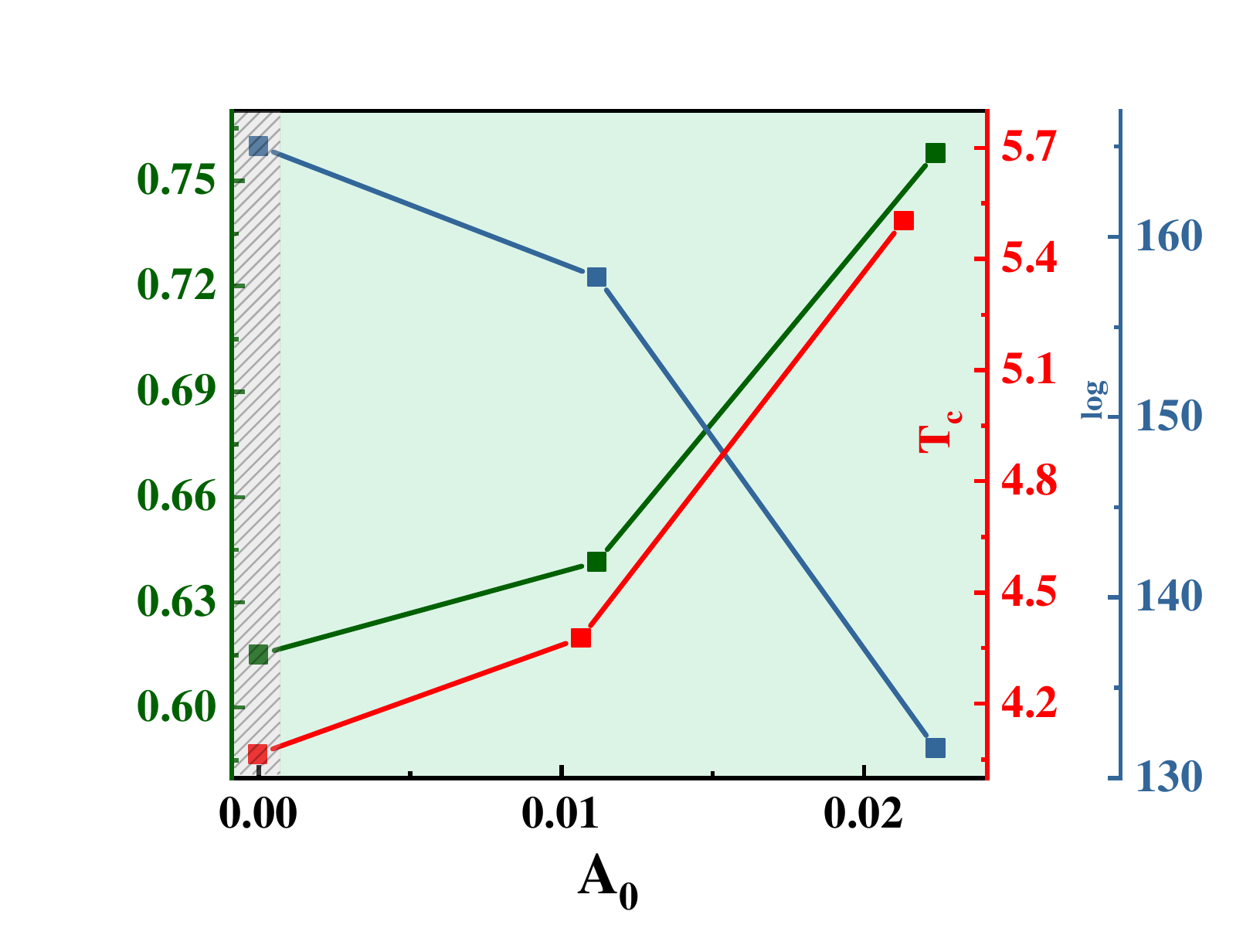}
  \caption{Cavity dependence of the electron--phonon coupling constant $\lambda$ (green), the logarithmic average phonon frequency $\omega_{\log}$ (blue), and the superconducting transition temperature $T_c$ (red) as a function of the cavity-field amplitude $A_0$. With increasing $A_0$, $\lambda$ increases while $\omega_{\log}$ decreases due to the redistribution of spectral weight toward lower frequencies. The resulting enhancement of $T_c$ indicates that the increase in $\lambda$ dominates over the reduction in $\omega_{\log}$.}
    \label{Fig.S3}
\end{figure}

\begin{figure}
  \centering
\includegraphics[scale=0.8]{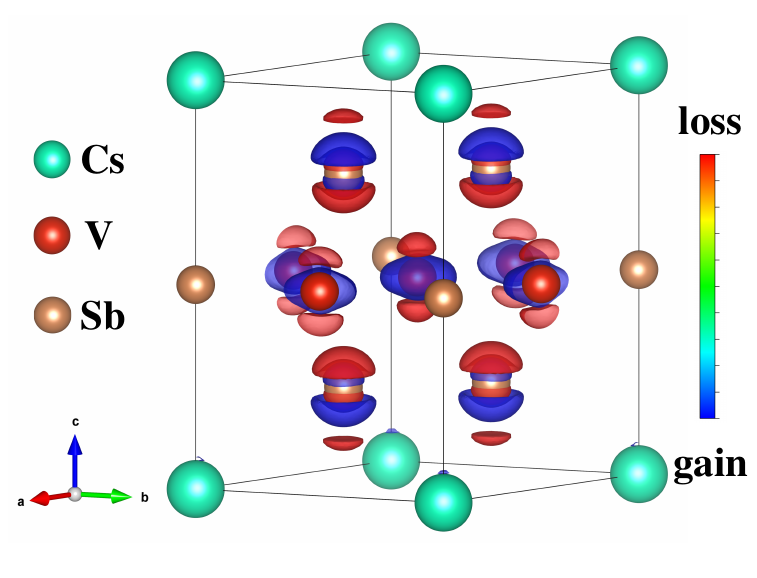}
  \caption{Three-dimensional isosurface plot of the cavity-induced charge-density difference in CsV$_3$Sb$_5$. The charge-density difference is defined as $\Delta \rho(\mathbf{r})=\rho_{A_0}(\mathbf{r})-\rho_0(\mathbf{r})$, where $\rho_{A_0}$ and $\rho_0$ are the charge densities with and without cavity coupling, respectively. Red and blue isosurfaces represent charge accumulation and depletion, respectively. The induced redistribution is concentrated around the V kagome layer and the surrounding Sb coordination environment, and shows a clear out-of-plane dipolar pattern along the cavity-polarization direction. This real-space charge response supports the interpretation that the $z$-polarized cavity acts as a polarization-selective perturbation to the CDW-related lattice dynamics.}
    \label{Fig.S4}
\end{figure}

\begin{table*}
\caption{Optimized lattice parameters and fractional atomic coordinates of CsV$_3$Sb$_5$ outside the cavity from 0 to 3 GPa. The structures are described in the conventional hexagonal cell with $P6/mmm$ symmetry (space group No.~191), and only symmetry-inequivalent atomic positions are listed.}
\label{tab:lattice_atomic_outside_cavity}
\begin{ruledtabular}
\begin{tabular}{cccccccc}
Pressure & $a$ & $c$ & $V$ & Cs & V & Sb1 & Sb2 \\
(GPa) & (\AA) & (\AA) & (\AA$^3$) & $(x,y,z)$ & $(x,y,z)$ & $(x,y,z)$ & $(x,y,z)$ \\
\hline
0 &
5.4538 &
9.3296 &
240.3184 &
$(0,0,0)$ &
$(1/2,0,1/2)$ &
$(0,0,1/2)$ &
$(1/3,2/3,0.2570)$ \\
1 &
5.4399 &
8.9937 &
230.4932 &
$(0,0,0)$ &
$(1/2,0,1/2)$ &
$(0,0,1/2)$ &
$(1/3,2/3,0.2488)$ \\
1.5 &
5.4327 &
8.8691 &
226.6960 &
$(0,0,0)$ &
$(1/2,0,1/2)$ &
$(0,0,1/2)$ &
$(1/3,2/3,0.2456)$ \\
2 &
5.4258 &
8.7600 &
223.3430 &
$(0,0,0)$ &
$(1/2,0,1/2)$ &
$(0,0,1/2)$ &
$(1/3,2/3,0.2428$ \\
3 &
5.4125 &
8.5822 &
217.7342 &
$(0,0,0)$ &
$(1/2,0,1/2)$ &
$(0,0,1/2)$ &
$(1/3,2/3,0.2381)$ \\
\end{tabular}
\end{ruledtabular}
\end{table*}


\bibliographystyle{apsrev4-2}

\clearpage
\bibliography{SI}  